\documentclass[aps,prd,twocolumn,superscriptaddress]{revtex4-1}
\usepackage{graphicx}
\usepackage{dcolumn}
\usepackage{bm}

\def\bea {\begin{eqnarray}}
\def\eea {\end{eqnarray}}

\def\be {\begin{equation}}
\def\ee {\end{equation}}

\begin{document}

\title{Indication of change of phase in high-multiplicity proton-proton events at LHC in String Percolation Model}
\author{I. Bautista}
\email{irais.buatista@fcfmbuap.mx}
\medskip
\author{A. Fernandez T\'ellez}
\affiliation{Facultad de Ciencias F\'isico Matem\'aticas, Benem\'erita Universidad Aut\'onoma de Puebla, 1152, M\'exico}
\author{Premomoy Ghosh}
\affiliation{Variable Energy Cyclotron Centre, 1/AF Bidhannagar, Kolkata 700 064, India}
\date{\today}

\begin{abstract}
We analyze high multiplicity proton-proton ($pp$) collision data in the framework of the String Percolation Model (SPM) that has been successful in describing 
several phenomena of multiparticle production, including the signatures of recent discovery of strongly interacting partonic matter, the Quark Gluon 
Plasma (QGP), in relativistic heavy-ion collisions.  Our study in terms of the ratio of shear viscosity and entropy density ($\eta/s$) and the (LQCD) predicted signature 
of QCD change of phase, in terms of effective number of degrees of freedom ($\epsilon /T^4$), reiterates the possibility of strongly interacting collective medium in these 
events. 
\end{abstract}


\maketitle

The Quark-Gluon Plasma (QGP) \cite{ref01}, an exotic state of matter of partonic degrees of freedom, is predicted \cite{ref02} by quantum chromodynamics (QCD) the theory of strong interaction. Experiments with relativistically colliding heavy nuclei at the Relativistic Heavy Ion Collider (RHIC) at Brookhaven National Laboratory
revealed features \cite {ref03,ref04,ref05,ref06}, expected from QGP-like dense partonic medium. Relativistic hydrodynamics helped in 
identifying the collective nature of the medium. Unpredictably, however, the ratio of the shear viscosity and the entropy density ($\eta /s$), the measure of fluidity, at the 
RHIC heavy-ion collisions reached a value \cite {ref07, ref08} close to that for a perfect fluid, indicating formation of a strongly interacting medium, termed as sQGP. Recent 
heavy-ion data from the Large Hadron Collider (LHC) at CERN with heavier nuclei and at higher energy corroborates \cite {ref09} most of the RHIC findings. 
In this scenario of the QGP study, an unexpected feature, namely the "ridge" in the long range near side angular correlation, in distinct class of "high multiplicity" events 
of proton-proton collisions \cite {ref10} at $\sqrt {s} = $ 7 TeV at LHC has triggered revival of an old school of thought  \cite {ref11, ref12, ref13, ref14} of the possibility of 
formation of a collective medium in $pp$ collisions also. Several of subsequent theoretical and phenomenological studies \cite {ref15, ref16, ref16a,ref16b,ref17, 
ref17a, ref17b,ref18,ref19}, in different approaches, endorse the possibility, indicating the need for further investigations in understanding the high-multiplicity $pp$ events 
vis-a-vis the QGP. 

In this article, we address the issue of collectivity in high multiplicity $pp$ events in the framework of the String Percolation Model, that has successfully explained the collectivity 
and the change of phase in nucleus-nucleus collisions \cite{ref20, ref21a, ref21b}. Also SPM 
describes the centre-of-mass energy dependence of mid-rapidity multiplicity \cite {ref21a} and the pseudorapidity distributions \cite {ref21b} of produced charged particles in 
$pp$ collisions, for the entire range of energy, available so far. In the SPM, the sources of multiparticle productions are the color strings between the colliding partons. The
stretched strings between the partons decay into new pairs of partons and so new strings are formed. Subsequently, particles are produced from interaction of partons by the Schwinger Mechanism. With the increasing energy of collision and / or the size of the colliding system, the density of the string increases and they start to overlap forming clusters. The overlapped strings start interacting. 
At a certain critical density, the strings start percolating through one another, forming a macroscopic cluster of strings - a geometrical phase transition takes place. The cluster 
of percolated color strings is considered to be equivalent to the de-confined partonic  state of matter \cite{ref20}. In fact, there has been considerable progress in the SPM in establishing connection \cite{ref20} between the percolation phase transition of color strings and the QCD phase transition in heavy-ion collisions. 

One of the cardinal parameters in SPM is the transverse impact parameter density of strings, $\zeta^{t}$. For $pp$ collisions, \cite{ref20} one can write 
$\zeta^{t}\equiv(\frac{r_{0}}{R_{p}})^{2}\bar{N}^{s}$ where $r_{0}$ is $\simeq 0.25$ fm the single string transverse size, $R_{p}\simeq 1$ fm is the proton transverse size 
and $\bar{N}^{s}$ is the average number of single strings.  For values of $\zeta^{t}$ above the critical value for the 2-dimensional percolation, $\zeta^{t}_{c}$ ($\simeq 1.2- 1.5$, depending on the profile function homogeneous or Wood Saxon type), $\zeta^{t}\gtrsim \zeta^{t}_{c}$, one observes the formation of long strings due to fusion and stretching 
between the colliding nucleons. Other important quantity in SPM is the Color Suppression Factor, F($\zeta^{t}$), which is related to the particle 
density $dN/dy$ and the number ($\overline{N}^{s}$) of strings as:
\begin{equation}
\frac{dN}{dy}=\kappa F(\zeta^{t})\bar{N}^{s}
\end{equation}
where $\kappa$ is a normalization factor $\sim .63$  \cite{ref21a} and
$F(\zeta^{t})\equiv \sqrt{\frac{1-e^{-\zeta^{t}}}{\zeta^{t}}}$ slows down the rate of increase in particle density with energy and with the number of strings. 

For $pp$ collisions one can approximately write 
\begin{equation}
N_{p}^{s}=2+4(\frac{r_{0}}{R})^{2} (\frac{\sqrt{s}}{m_{p}})^{2\lambda}
\end{equation}
with $m_{p}$ the mass of the proton and $\lambda$ a constant parameter $\simeq.201$ \cite{ref21a}.

By measuring \cite{ref33} the $d\langle N_{ch} \rangle /d\eta$ - dependent identified particle spectra  from $pp$  collisions at $\sqrt {s}$ = 0.9, 2.76 and 7 TeV, 
the CMS experiment at the LHC provides the unique opportunity for somewhat  "centrality" dependent study in $pp$ collisions. To determine the corresponding value 
of the string density, we use the invariant transverse momentum spectra given by a power law:
\begin{equation}
\frac{1}{N}\frac{d^{2}N}{dp_{T}^{2}}=\frac{(\alpha-1)(\alpha-2)}{2\pi p_{0}^{2}}\frac{p_{0}^{\alpha}}{[p_{0}+p_{T}]^{\alpha}}
\end{equation}
where $p_{0}$ and $\alpha$ are energy dependent parameters. 
The total multiplicity is obtained by the mean over all the clusters configurations
$ \mu=\left \langle \sum_{i=1}^{M} \sqrt{\frac{n_{i}S_{i}}{S_{1}}} \right \rangle \mu_{1}$ where $M$ is the total number of clusters in a event and $n_{i}$ the number of strings that form the $i$ cluster. 
Considering only the clusters which contributes to the central region the general formula: 
\begin{equation}
\frac{dN_{ch}}{d\eta}\mid_{\eta=0}=\frac{\left \langle \sum_{i=1}^{M} \sqrt{\frac{n_{i}S_{i}}{S_{1}}} \right \rangle}{ \left \langle \sum_{i=1}^{M} \sqrt{\frac{n_{i}S_{i}}{S_{1}}} \right \rangle_{pp}} \frac{dN_{ch}}{d\eta} \mid_{\eta=0}^{pp}
\end{equation}
with  $N_{p}^{s}=\left \langle \sum_{i=1}^{M} \sqrt{\frac{n_{i}S_{i}}{S_{1}}}\right \rangle \mid_{pp}$
and similarly 
\begin{equation}
\langle p_{T}^{2} \rangle =\frac{\left \langle \frac{N}{\sqrt{\frac{nS_{n}}{S_{1}}}}\right \rangle}{\left \langle\frac{N}{\sqrt{\frac{n S_{n}}{S_{1}}}} \right \rangle_{pp}} \left \langle p_{T}^{2} \right \rangle_{pp}
\end{equation}
By using equation (4) and (5) and (3) we can describe the general equation that relates the high multiplicity (with string density $\zeta_{HM}$) and min bias distributions as:

\begin{equation}
\frac{1}{N}\frac{d^{2}N_{ch}}{d\eta dp_{T}}|_{\eta=0}=a
\frac{(p_{0}b)^{\alpha-2}}{(p_{T}+p_{0}b)^{\alpha-1}}
\end{equation}
with $a=\frac{\left \langle \sum_{i=1}^{M}\sqrt{\frac{n S_{i}}{S_{1}}} \right \rangle}{\left \langle \sum_{i=1}^{M}\sqrt{\frac{n S_{i}}{S_{1}}}\right \rangle_{pp}}\frac{dN}{d\eta}|_{\eta=0}^{pp}
 \frac{(\alpha-2)}{2\pi}$ and $b=\left (\left \langle \frac{N}{\sum_{i=1}^{M}\sqrt{\frac{ni S_{i}}{S_{1}}}}\right \rangle / \left \langle\frac{N}{\sum_{i=1}^{M}\sqrt{\frac{n_{i}S_{i}}{S_{1}}}} \right \rangle_{pp}\right )^{1/2}$,
by applying the thermodynamic limit with a vectorial color sum $b\rightarrow \sqrt{F(\zeta)/F(\zeta_{HM})}$. So, finally one gets 
\begin{equation}
\frac{1}{N} \frac{d^{2}N}{d\eta dp_{T}}=\frac{a
(p_{0}\sqrt{\frac{F(\zeta_{pp})}{F(\zeta_{HM})}})^{\alpha-2}
}{[p_{0}\sqrt{\frac{F(\zeta_{pp})}{F(\zeta_{HM})}}+p_{T}]^{\alpha-1}}
\end{equation}
Note that for $pp$ collisions minimum (7) reduces to 
\begin{equation}
\frac{1}{N} \frac{d^{2}N}{d\eta dp_{T}}=\frac{a p_{0}^{\alpha-2}
}{[p_{0}+p_{T}]^{\alpha-1}},
\end{equation}
To obtain $a$, $p_{0}$, $\alpha$ we perform a fit to the transverse momentum distributions of charged particles from minimum bias $pp$ events at the energies $\sqrt{s}=900$ GeV, 
2.76 TeV, 7 TeV \cite{ref33} with equation (8) (see Table 1). To determine the values of $\zeta_{MH}$ equation (7) was fit to (presented in Figure 1) the measured transverse momentum 
distributions \cite{ref33} with the corresponding values from Table 1. The fit is restricted to $p_{T} > .4$ GeV/c to avoid the effect 
of resonance decays. For $dN/d\eta$,  we have 
taken into account the kinematics cuts, by scaling the measured $\langle N_{track} \rangle$ with the corresponding factor of ($1/4.8$) corresponding to the $|\eta|<2.4$ 
range and the factor ($1.6$ ) as in \cite{ref35a} corresponding to the $p_{T}$ cut range. 

\begin{table}[h]
\scalebox{.9}{
\centering
\begin{tabular}{cccc}
\hline
\hline
$\sqrt{s}$  (TeV) & a & $p_{0}$ & $\alpha$ \\

.9 & 23.29 $\pm$ 4.48 & 1.82 $\pm$ .54& 9.40 $\pm$ 1.80 \\

2.76 & 22.48$\pm$ 4.20 & 1.54 $\pm$ .46 & 7.94$\pm$ 1.41\\

7 & 33.12 $\pm$ 9.30 & 2.32 $\pm$ .88 & 9.78 $\pm$ 2.53\\
\hline
\hline
\end{tabular}
}
\caption {Parameters of the transverse momentum distribution (9) in $pp$ collisions.}
\end{table}
\begin{figure}[ht]
 \includegraphics[width=.9\columnwidth]{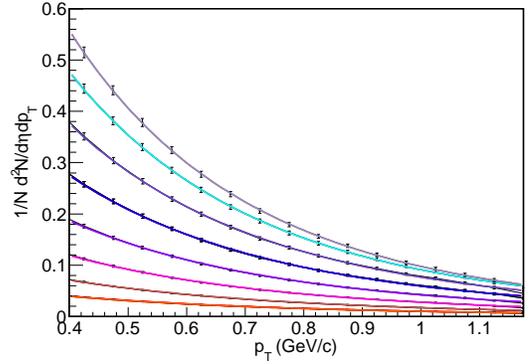}%
 \caption{Fits to the transverse momentum distribution for energies $\sqrt{s}=$ 7 TeV in $p-p$ collisions for different multiplicity classes from $N_{track}=131$ grey line to $N_{track}=40$ orange line. Data taken from reference \cite{ref33}.}
  \end{figure}

 Figure 2 show the dependence of the obtained color reduction factor $F(\zeta^{t})$ with the corresponding $dn/d \eta$ respectively at different energies, it indicates that at the same multiplicity class the string density decreases for higher energies.
\begin{figure}[ht]
 \includegraphics[width=.9\columnwidth]{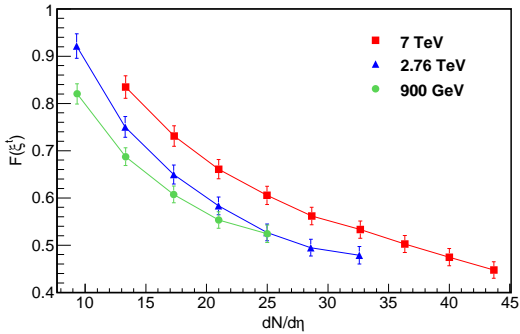}%
 \caption{Color reduction factor at high multiplicities for different energies}
 \end{figure}

The Schwinger mechanism \cite{ref21c} for massless particles is given by the expression 
\begin{equation}
\frac{dN}{dp_{T}}\sim e^{-\sqrt{2 F(\zeta^{t})} \frac{p_{T}}{\langle p_{T} \rangle _{1}}},
\end{equation}
 which can be related to the average value of the string tension
$\langle x^{2} \rangle =\pi \langle p_{T}^{2}  \rangle_{1} /F(\zeta) $ \cite{ref29b}, 
this value fluctuates around its mean value because the chromoelectric field is not constant, the fluctuations of the chromo electric field strength lead to a Gaussian distribution of the string tension that transform it into a thermal distribution, where the temperature is given by the relation \cite{ref19}
\begin{equation}
T(\zeta^{t})=\sqrt{\frac{\langle p_{T}^{2} \rangle _{1}}{2F(\zeta^{t})}}
\end{equation}

\begin{figure}[ht]
 \includegraphics[width=.9\columnwidth]{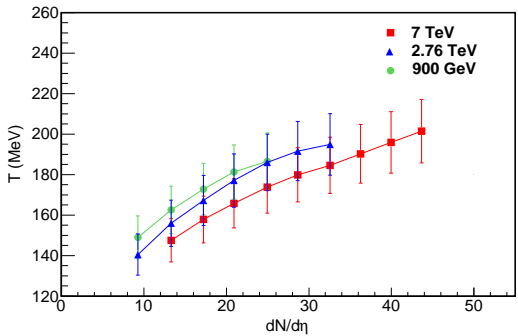}%
 \caption{Efective temperature vs  $dn/d\eta$.}
 \end{figure}

We consider that the experimentally determined chemical freeze out temperature is a good measure of the phase transition temperature $T_{c}$.
We calculate the effective temperature, $T$, from the equation (10), for each multiplicity class for a critical density $\zeta_{c}=1.2$ and at the critical temperature 
$T_{c}=154 \pm 9$ MeV, as obtained by the latest LQCD results from the HotLQCD collaboration \cite{ref34a}, with the corresponding $\langle p_{T} \rangle_{1}\sim 190.25 \pm 11.12 $ MeV/c consistent with the measured of direct photon enhanced measured \cite{ref29b}.
Figure 3 shows the increase of calculated $T$ with the increase of the multiplicity $dn/d\eta$.

 \begin{figure}[ht]
\includegraphics[width=.9\columnwidth]{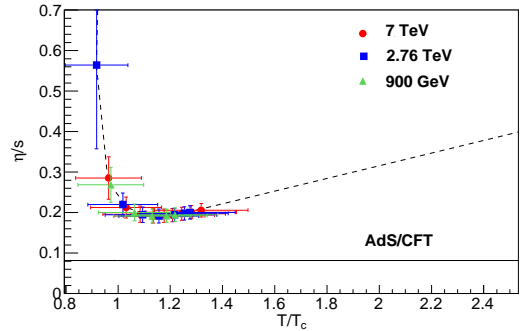} 
\caption{Shear viscosity over entropy ratio for 7 TeV high multiplicity classes corresponding to $N_{track}=40$ to $N_{track}=131$, with the $T_{c}=154\pm9$ MeV.}
\end{figure}

In terms of the effective temperature one can study some useful quantities as the ratio of the shear viscosity over entropy density, mean transverse momentum and the 
ratio of energy density over $T^{4}$ as it shown in the following lines:

In the relativistic kinetic theory as
$\eta/s \simeq \frac{T\lambda_{fp}}{5}$
where $\lambda_{mfp}$ is the mean free path $\sim\frac{1}{n\sigma_{tr}}$, n is the density of the effective number of sources per unit volume and $\sigma_{tr}$ is the 
transport cross section, $n=\frac{N_{sources}}{S_{N}L} $. It is considered that $\frac{N_{sources}}{S_{N}L} \sigma_{tr} = (1-e^{-\zeta^{t}})/L$
considering $L=1 fm$ the longitudinal extension of the source one can give the relation $\eta/s$ in terms of $\zeta^{t}$ \cite{ref20}, as:
\begin{equation}
\frac{\eta}{s}=\frac{TL}{5(1-e^{-\zeta^{t}})}
\end{equation}

The $\eta/s$ as a function of $T/T_{c}$ as obtained from the SPM formalism for the $pp$ data \cite{ref33} is shown in Figure 4. The plot shows similar features 
as has been exhibited by the heavy-ion data \cite{ref07} \cite{ref29b}.

\begin{figure}[ht]
\includegraphics[width=.9\columnwidth]{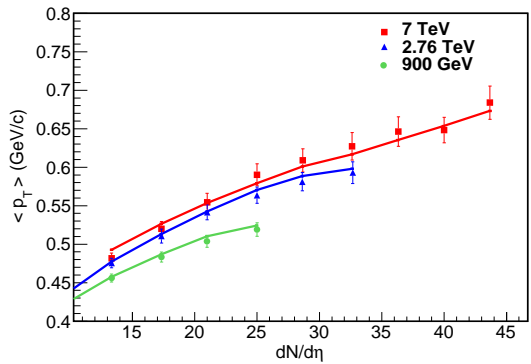}%
\caption{Mean transverse momentum at different energies data from \cite{ref33}}
\end{figure}

The evolution of the mean transverse momentum can also described from (6) as an inverse function of the color reduction factor
$ \langle p_{T} \rangle \simeq p_{0}\sqrt{F(\zeta)/F(\zeta_{HM})}\frac{2}{(\alpha-3)}$ The Figure 5 clearly shows that the SPM model describes the measured \cite{ref33}
$dn_{ch}/d\eta$ -dependence of $ \langle p_{T} \rangle$.

In SPM the energy density can be determined by a temperature at string level, by using the energy density from Bjorken
$\epsilon_{i}=\frac{3}{2}\frac{\frac{dN_{c}}{d y}\langle p_{T} \rangle}{S_{N}\tau_{pro}} $,
 with the nuclear overlap area $S_{n}$, and taking the production time for a boson gluon ($\tau_{pro}$) \cite{ref20} as the 
 propagation time of the parton given in fermi ($\tau=2.405\hbar/<m_{t}>$). The results are shown in Figure 6.

\begin{figure}[ht]
\includegraphics[width=.8\columnwidth]{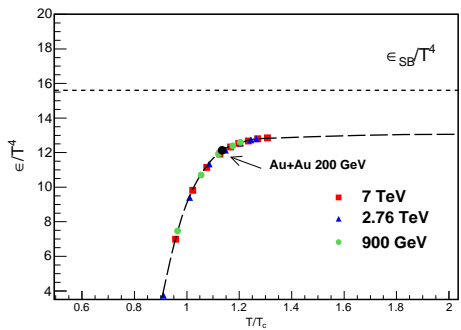}
\caption{Energy density at different energies}
\end{figure}

The $dn_{ch}/d\eta$-dependence $\langle p_{T} \rangle$ of the $pp$ data at the LHC energies is well described by the String Percolation Model. Besides, in 
terms of the ratio of shear viscosity and entropy density ($\eta/s$) and the (LQCD) predicted signature of QCD change of phase, in terms of effective number of 
degrees of freedom ($\epsilon /T^4$), the model gives a clear indication of the phase transition for the measured high multiplicity $pp$ events. Of course, there 
have been several other models to describe the unexpected behavior of high multiplicity $pp$ events. Nevertheless, till one can analyze expected high statistics 
of high multiplicity $pp$ events at $\sqrt s$ = 13 TeV, in the Run-2 of LHC, the SPM provides one of the possible explanations for the observed unexpected
features of the $pp$ data at LHC energies.

\section{Acknowledgments}
I. Bautista was supported by CONACYT grant. We would like to thank C. Pajares and B. K. Srivastava for fruitful discussions.


\begin{thebibliography}{99}
\bibitem {ref01} Collins, J. C. and Perry, M. J.,Phys. Rev. Lett. {\bf 34},1353-1356(1975).
\bibitem{ref02} Shuryak, E., Phys. Reports {\bf 61}, 71-158 (1980).
\bibitem{ref03} Arsene, I. et al., BRAHMS Collaboration, Nucl. Phys.
{\bf A757}, 1-27 (2005).
\bibitem{ref04} Back, B. B. et al., PHOBOS Collaboration. Nucl. Phys.
{\bf A757}, 28 - 101 (2005).
\bibitem{ref05}Adams, J. et al., STAR Collaboration, Nucl. Phys.
{\bf A757}, 102-183 (2005).
\bibitem{ref06} Adcox, K. et al., PHENIX Collaboration, Nucl. Phys.
{\bf A757}, 184-283 (2005).
\bibitem{ref07} Hirano, T. and Gyulassy, M., Nucl. Phys. {\bf A769}, 71 (2006).
\bibitem{ref08} Csernai, L.P. Kapusta, J. I. and McLerran, L.,  M. Phys. Rev. Lett. {\bf 97}, 152303 (2006).
\bibitem{ref09} Muller, B, Schukraft, J. and Wyslouch, B., Annu. Rev. Nucl. Part. Sci. {\bf62}, 361 (2012).
\bibitem{ref10} Khachatryan, V. et al., CMS Collaboration, J. High Energy Phys. {bf 09 }, 091 (2010).
\bibitem{ref11} Van Hove, L., Phys. Lett.{\bf B118}, 138 (1982).
\bibitem{ref12} Bjorken, J. D. Fermilab Report No. FERMILAB-PUB-82-059-THY, (1982).
\bibitem{ref13} Alexopoulos. T, et al.,  Phys. Lett.{\bf B528}, 43 - 48 (2002).
\bibitem{ref14} Levai, P. and Muller, B. Phys. Rev. Lett. {\bf 67}, 1519 (1991).
\bibitem{ref15} Bozek, P. Eur. Phys. J.{\bf C71}, 1530 (2011).
\bibitem{ref16} Werner, K., Karpenko, L. and Pierog, T. Phys. Rev. Lett. {\bf 106}, 122004 (2011).
\bibitem{ref16a} Shuryak, E. and Zahed, I. Phys. Rev.{\bf C88}, 044915 (2013). 
\bibitem{ref16b} Ghosh, P., Muhuri, S., Nayak, J. and Varma, R., J. Phys. G {\bf 41} 035106 (2014).
\bibitem{ref17} Ghosh, P. and Muhuri, S., arXiv:1406.5811 [hep-ph].
\bibitem{ref17a} I. Bautista, Proceedings of ICPAQGP2015, 2-6 February 2015 (to be published).
\bibitem{ref17b} L. G. Gutay, A. S. Hirsch, C. Pajares, R. P. Scharenberg and B. K. Srivastava, arXiv:1504.08270 [nucl-ex].
\bibitem{ref18} A. Ortiz Velasquez, P. Christiansen, E. Cuautle Flores, I. Maldonado Cervantes and G. Paic,  Phys.\ Rev.\ Lett.\  {\bf 111} (2013) 4,  042001
\bibitem{ref19} A. V. Ortiz,  Nucl.\ Phys.\ A {\bf 943} (2015) 9.
\bibitem{ref20}  Braun, M. A. Dias de Deus, J. Hirsch, A. S. Pajares, C. Scharenberg, R. P. and Srivastava, B. K.,  arXiv:1501.01524 [nucl-th].
\bibitem{ref21a} Bautista, I. Milhano, J. G. Pajares, C. and Dias de Deus, J. Phys. Lett. {\bf B715} 230 (2012).
\bibitem{ref21b} Bautista, I. Dias de Deus, J. and Pajares, C. Phys. Rev. {\bf C86} 034909 (2012).
\bibitem{ref21c} J. S. Schwinger, Phys. Rev.  {\bf 128} (1962) 2425.
\bibitem{ref29b} R. P. Scharenberg, B. K. Srivastava, A. S. Hirsch, Eur. Phys. J. C 71 (2011) 1510.
\bibitem{ref33} S. Chatrchyan {\it et al.}  [CMS Collaboration], Eur. Phys. J. C {\bf 72} (2012) 2164.
\bibitem{ref35a} E.~Shuryak and I.~Zahed, Phys.\ Rev.\ C {\bf 88} (2013) 4,  044915.
\bibitem{ref34a} HotQCD: Phys. Rev. D85, 054503 (2012).
\bibitem{ref36} N. Cartiglia, arXiv:1305.6131 [hep-ex].
\end{thebibliography}
\end{document}